\documentclass[fleqn,10pt]{wlscirep}

\usepackage{physics}
\usepackage[english]{babel}
\usepackage[utf8]{inputenc}
\usepackage[T1]{fontenc}
\usepackage{amssymb}
\usepackage{amsmath}
\graphicspath{
    {./Figures/}
    {./}
}

\usepackage{bm}
\usepackage{hyperref}
\hypersetup{
    colorlinks,%
    citecolor=blue,%
    filecolor=blue,%
    linkcolor=blue,%
    urlcolor=blue
}

\usepackage{xcolor}

\usepackage{xspace}
\providecommand{\tqwbhz}{3$\times$BHZ\xspace}

\usepackage[normalem]{ulem}

\title{Engineering topological phases in triple HgTe/CdTe quantum wells}

\author[1]{G. J. Ferreira}
\author[2]{D. R. Candido}
\author[3]{F. G. G. Hernandez}
\author[3]{G. M. Gusev}
\author[4]{E. B. Olshanetsky}
\author[4]{N. N. Mikhailov}
\author[4]{S. A. Dvoretsky}

\affil[1]{Instituto de Física, Universidade Federal de Uberlândia, Uberlândia, MG 38400-902, Brazil}
\affil[2]{Department of Physics and Astronomy, University of Iowa, Iowa City, Iowa 52242, USA}
\affil[3]{Instituto de Física, Universidade de São Paulo, São Paulo, São Paulo 05508-090, Brazil}
\affil[4]{Institute of Semiconductor Physics, Novosibirsk 630090, Russia}

\begin{document}

\date{\today}

\begin{abstract}
Quantum wells formed by layers of HgTe between Hg$_{1-x}$Cd$_x$Te barriers lead to two-dimensional (2D) topological insulators, as predicted by the BHZ model. Here, we theoretically and experimentally investigate the characteristics of triple HgTe quantum wells. We describe such heterostructure with a three dimensional $8\times 8$ Kane model, and use its eigenstates to derive an effective 2D Hamiltonian for the system. From these we obtain a phase diagram as a function of the well and barrier widths and we identify the different topological phases composed by zero, one, two, and three sets of edge states hybridized along the quantum wells. The phase transitions are characterized by a change of the spin Chern numbers and their corresponding band inversions. Complementary, transport measurements are experimentally investigated on a sample close to the transition line between the phases with one and two sets of edges states. Accordingly, for this sample we predict a gapless spectrum with low energy bulk conduction bands given by one parabolic and one Dirac band, and with edge states immersed in the bulk valance bands. Consequently, we show that under these conditions, local and non-local transport measurements are inconclusive to  characterize a sole edge state conductivity due to bulk conductivity. On the other hand, Shubnikov-de Haas (SdH) oscillations show an excellent agreement with our theory. Particularly, we show that the measured SdH oscillation frequencies agrees with our model and show clear signatures of the coexistence of a parabolic and Dirac bands.
\end{abstract}
\flushbottom
\maketitle
\thispagestyle{empty}

\section*{Introduction}

The discovery of two and three dimensional (2D and 3D) topological insulators (TIs), also known as quantum spin hall (QSH) insulators, strongly impacted the field of quantum materials due to their interesting fundamental properties and technological applications \cite{kane1, kane2, Bernevig2006QSHE, Bernevig2006BHZ, HasanRMP, ZhangRMP, Qi_2010, Kvon_2020}. They constitute a peculiar class of materials, characterized by an insulating bulk dispersion and gapless topological helical surface or edge states that are shown to be protected against back-scattering by the time reversal symmetry. The first theoretical prediction for a TI was proposed by Haldane \cite{Haldane1988}, and it is built upon a ``2D graphite'' spinless toy model. Although this proposal lacked physical justifications at that time, it was later theoretically realized in spinful graphene \cite{kane2} despite its spin-orbit gap being too small \cite{min} to make them experimentally realizable. In fact, whereas a broad variety of semiconductor-based  materials can host topological helical states\cite{catalogue}, the first experimental indication of edge channels conductivity was reported in HgTe/CdTe composite quantum wells (QWs) \cite{Konig2007HgTe, Roth}, following the theoretical prediction of the BHZ model \cite{Bernevig2006QSHE, Bernevig2006BHZ} (named after its authors Bernevig, Hughes, and Zhang). Additionally, the 2D Dirac-like band structure of 3D TIs have been observed in ARPES measurements \cite{Hsieh_2008, Hsieh_2009, Xia_2009, Chen_2010, Brune2011, Neupane2012}.

Particularly in 2D TIs, the energy ordering of electron-like and hole-like QW eigenstates leads to the topological phase transition from a trivial insulator to a TI. This is shown to be controlled by the QW width\cite{Bernevig2006QSHE, Bernevig2006BHZ}, electric field \cite{Li_2009, Rothe_2010, Ezawa_2013, Krishtopenko2016DW, Baradaran_2020}, strain \cite{Krishtopenko2016Phases, Mahler_2021}, and temperature\cite{Krishtopenko2016Phases, Kadykov2018Temp}. The resulting 1D helical channels at the edges lead to quantized conductance and nonlocal edge transport, which has been observed for sufficiently short distances between measurement probes \cite{Gusev_2019, Hsu_2021}. In all cases, the quantized conductance of HgTe-base QWs shows significant fluctuations, with experimental values different from the ones predicted theoretically through Landauer-Buttiker formalism. The deviation from the theoretical prediction has been attributed to many different effects, including disorder\cite{Baum2015CoexistDisorder, Krishtopenko2020Disorder}, charge puddles \cite{Vayrynen2013Puddles, Lunczer2019}, and many sources of inelastic scattering \cite{Grabecki2013}. Despite the deviation from theoretical results, recent improvements in sample growth has lead to measurements closer to the predicted quantization \cite{Bendias_2018, Lunczer2019}.

Recently, it has been proposed that the control of the layer localization of topological states in bilayer graphene\cite{Pelc2015, Jaskolski_2016} and TIs\cite{Crasto_de_Lima_2019} could be used to design ``\textit{layertronic}'' devices, where this additional degree of freedom could be used together with the spin, valley, and charge to build novel devices. For HgTe-based 2D TIs, the multilayer character arises from multiple coupled QWs. In the bilayer case, \textit{i.e.} double HgTe QWs (DQWs), an intuitive 2D model for coupled QWs has been introduced by Refs.~[\citen{Michetti2012DW, Michetti_2013}], with Ref.~[\citen{Krishtopenko2016DW}] showing that a variety of topological phases can be obtained depending upon the QW geometrical parameters. Additionally, it has been recently proposed\cite{Liu2021} that these DQWs could host second-order topological insulators with excitonic nodal phases that support ﬂat band edge states, which could lead to superconductivity \cite{aoki}. Experimentally, signatures of the conductance quantization, nonlocal transport and the Landau fan diagram of HgTe-based DQWs have been recently observed\cite{Gusev2020DW, Gusev2021DW}. Additionally, magnetically doped TI layers, coupled through insulating spacers, lead to a solid state realization of the Weyl semimetal\cite{burkov}, providing a platform to examine interesting topological features such as Fermi arc surface states and the chiral anomaly effect \cite{HasanRMP}. The Weyl semimetal can also be achieved in time-reversal invariant systems with broken inversion symmetry \cite{Wang_2017, Yan_2017}, which can be realized in multiple coupled HgTe-based QWs \cite{balents, Zyuzin2012}. Further advance on the physics of the multilayer Dirac fermions can be achieved in trilayer systems, \textit{i.e.} triple quantum wells (TQW). In contrast to the DQW case, the additional layer of the TQW allows for an interplay between the hybridization of the inner and outer layers, which can be controlled by its geometric parameters (wells and barrier widths), external electric fields and gates.

In this paper, we investigate the band structure and transport of triple HgTe/Hg$_{1-x}$Cd$_x$Te quantum wells, comparing theoretical transport predictions with experimental measurements of local and nonlocal resistivities, and Shubnikov–de Haas (SdH) oscillations. First, we obtain the band structures calculated with the $8\times 8$ Kane Hamiltonian, and determine the corresponding topological phase diagram as a function of the geometric TQW parameters (well and barrier widths). We find that electron-like (E) and hole-like (H) eigenstates are hybridized states spread throughout the three QWs as symmetric and anti-symmetric combinations. For the E-like bands the hybridization leads to large energy splitting between these bands, while the H-like bands remain nearly degenerate due to its larger effective mass. Additionally, we show that by increasing the QW widths, only crossings between E-like and H-like bands with equivalent envelope wavefunction profiles along the wells lead to topological phase transitions. Using their spin Chern number, we label four different phases of our system, namely, 0 (trivial), and I, II, III, referring to the number of pairs of edge states in each phase. Particularly, for phases I and II we show that bulk band structure is gapless and the edge states always coexist with the valence bands. Only for the phase III a bulk gap opens and the three edge states become isolated from the bulk bands within the gap region. 

Using the theoretical prediction of the topological phase diagram corresponding to our HgTe-TQW, we are able to predict that the experimental sample analyzed in this work lies in phase I, close to the transition towards phase II. Within this regime, we show that resistivity measurements do present signatures of nonlocal transport. However, these are strongly affected by bulk conductivity and scattering between different states, which leads to an inconclusive characterization of the total number of edge states conducting the current. On the other hand, the measurements of SdH oscillations shows a good agreement with the theoretically predicted SdH frequencies and also with the temperature dependence of the SdH oscillations. More specifically, we explained the SdH oscillation data as arising from an interplay of one linear and one parabolic conducting bands, which are shown to be in agreement with our band structure calculation. This analyzes is shown to be also in agreement with dependence of the peak nominal values and their spectral weight as a function of the Fermi energy.

\section*{Theoretical models}

In this section we introduce the theoretical models used to obtain the bulk energy bands, edge state energies, and all the corresponding wave-functions of our system. 

\begin{figure}[ht]
 \centering
 \includegraphics[width=\textwidth]{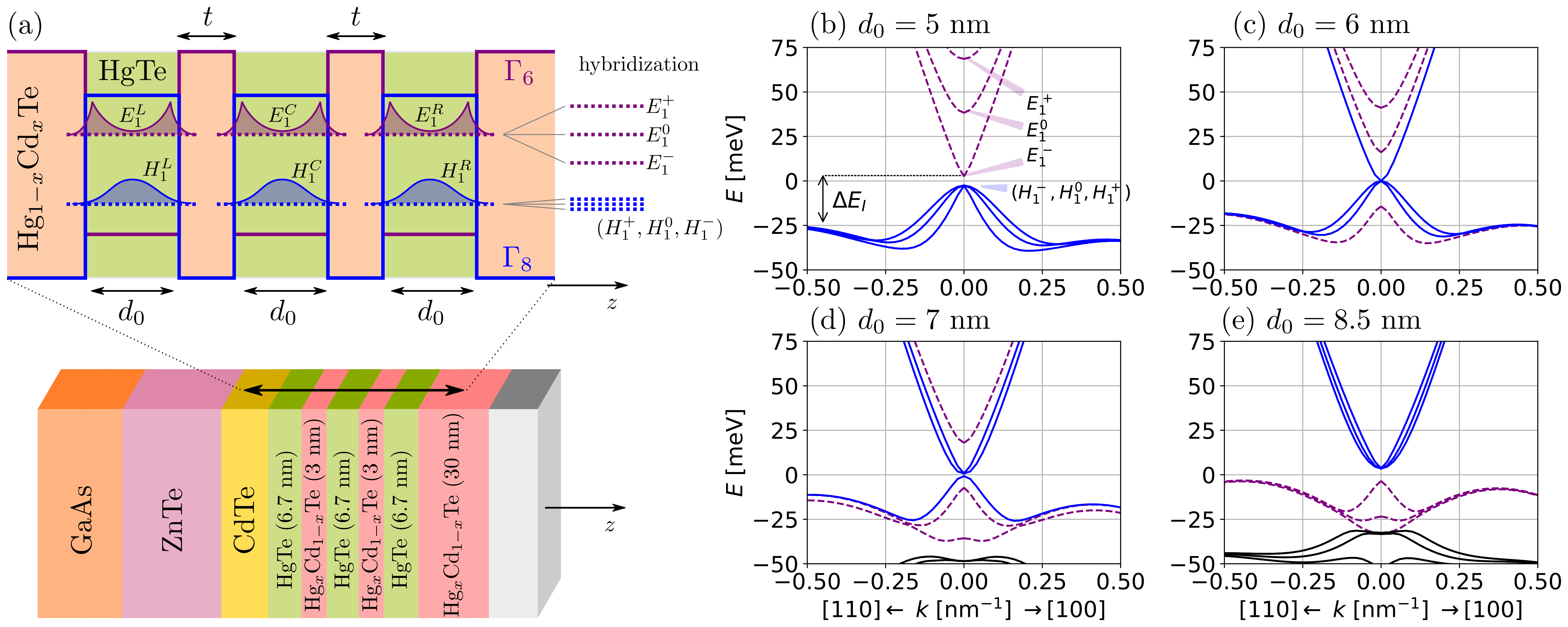}
 \caption{
 (a) Conduction and valence band edges of the triple quantum well schematically shown in the bottom. The widths $d_0$ of the HgTe wells and the thickness $t$ of the Hg$_{1-x}$Cd$_x$Te barriers ($x=0.3$) are indicated. Single well states located on each of left (L), central (C) and right (R) wells are illustrated in purple (electron-like) and blue (hole-like). As $t$ decreases their hybridization leads to a large energy splitting of the E-like states, while the H-like remain nearly degenerate due to the larger effective mass. (b,c,d) Band structures for $t=3$~nm and varying $d_0$. (b) For $d_0 = 5$~nm, in the trivial regime, all E bands are above the H bands. $\Delta E_I$ indicates the indirect band gap. Phase transitions occur as each E band crosses the H bands with increasing (c) $d_0 = 6$~nm, (d) $d_0 = 7$~nm, and (e) $d_0 = 8.5$~nm. For large $d_0$ (not shown) the indirect gap $\Delta E_I$ closes and the system becomes semi-metallic.
 }
 \label{fig:1}
\end{figure}

\subsection*{\texorpdfstring{$8\times 8$ Kane model}{8x8 Kane model}}

We consider quantum wells (QWs) made of HgTe confined by Hg$_{1-x}$Cd$_x$Te barriers with concentration $x = 0.3$, as shown in Fig.~\ref{fig:1}(a). Both HgTe and CdTe crystallize in the zindblende structure with low energy bands around the $\Gamma$ point ($\bm{k}=0$), which are well described by the $8\times 8$ Kane Hamiltonians\cite{winkler2003spin, voon2009kp} $H_{\rm Kane}$, with the corresponding
\{
$\ket{\Gamma_6, \pm \frac{1}{2}}$,
$\ket{\Gamma_7, \pm \frac{1}{2}}$,
$\ket{\Gamma_8, \pm \frac{1}{2}}$,
$\ket{\Gamma_8, \pm \frac{3}{2}}$
\} basis set. The CdTe bandstructure has a normal order, where $\Gamma_6$ is a S-type conduction band, $\Gamma_8$ and $\Gamma_7$ are the P-type valence bands corresponding to heavy and light holes ($\Gamma_8$) and the split-off band ($\Gamma_7$). In contrast, HgTe has the $\Gamma_6$ and $\Gamma_8$ bands inverted due to relativistic fine structure corrections (Darwin, spin-orbit, and mass-velocity terms), which ultimately allows for the QSH topological phase of single HgTe QWs \cite{Bernevig2006QSHE, Bernevig2006BHZ}. For heterostructures, one considers the Kane parameters to be position dependent, restores $\hbar\bm{k} \rightarrow \bm{p}$ as the momentum operator, and symmetrizes\cite{Bastard1988, voon2009kp, winkler2003spin} the Hamiltonian. Additionally, the growth of the heterostructure typically induces strain, which is considered under the Bir-Pikus Hamiltonian \cite{bir1974symmetry}. The resulting $8\times 8$ Kane Hamiltonian and the material parameters are shown in the Supplementary Material\cite{SupplementalMaterial}. Here, the theoretical model set the growth direction $z \parallel [001]$. Nevertheless, for small $d_0\lesssim 7.5$~nm we expect the results to be nearly equivalent for growth directions [001] and [013]\cite{Raichev2012Direction}. In the $(x,y)$ plane, the solutions are given by plane-waves, $\propto e^{i\bm{k}_\parallel \cdot \bm{r}}$, where $\bm{k}_\parallel = (k_x, k_y)$ is the in-plane momentum. To numerically diagonalize the $H_{\rm Kane}(z, p_z, \bm{k}_\parallel)$ we use the {\tt kwant} code \cite{Groth2014Kwant}, which provides an efficient interface to build and solve the numerical problem.

\subsection*{Effective 2D Hamiltonian for triple wells}
\label{sec:BHZx3}

To investigate the confinement of the subbands of our triple HgTe/CdTe quantum wells, their corresponding topological character and the characteristics of their edge states, we consider an \emph{effective} 2D Hamiltonian for our system. For single HgTe-quantum wells, this is achieved by the projection of the Hamiltonian into its $\bm{k}_\parallel=0$ eigenstates, which leads to the well known BHZ model \cite{Bernevig2006BHZ, Bernevig2006QSHE}. In contrast, for double HgTe-quantum wells there are two interesting approaches. First, similarly to the derivation of the BHZ Hamiltonian, in Refs.~[\citen{Krishtopenko2016DW, Krishtopenko2018Realistic}] the authors project the total Hamiltonian into the $\bm{k}_\parallel=0$ DQW eigenstates. Alternatively, in Ref.~[\citen{Michetti2012DW}] the authors project the total DQW Hamiltonian into the subbands of the single wells (left and right), and introduce tunneling parameters for the coupling between neighboring QWs. Here, for the case of triple QWs, we follow the approach from the latter, as it provides an intuitive perturbative picture of the coupling between quantum wells. This is illustrated schematically by Fig.~\ref{fig:1}(a). For easy reading, we keep here a notation similar to Ref.~[\citen{Michetti2012DW}], but we introduce the index $\nu=\{{L, C, R}\}$ to label the quantities of the individual left (L), central (C) and right (R) QWs. Accordingly, we define the subbands of the isolated QW as $\{\ket{H_{1\pm}^\nu}, \ket{E_{1\pm}^\nu}\}$, with $\pm$ labels referring to time-reversal partners. Assuming that tunnel coupling occurs only between neighboring QWs, it is immediate to extend the double QWs model\cite{Michetti2012DW} into the triple well case, which we label ``\tqwbhz'', and it reads as
\begin{align}
    \label{eq:H2D}
    H_{2D} &=
    \begin{pmatrix}
     H_L & V_{LC} & 0 \\
     V_{LC}^\dagger & H_C & V_{CR} \\
     0 & V_{CR}^\dagger & H_R
    \end{pmatrix}.
\end{align}
Here, the diagonal blocks of each layer, $H_\nu$, are given by a direct sum over Kramers partners $H_\nu = h_\nu(\bm{k}) \oplus h_\nu^*(-\bm{k})$, each composed by BHZ-like Hamiltonians
\begin{align}
    h_\nu(\bm{k}) =& \; 
    (C_\nu-D_\nu k^2) + A_\nu(\sigma_x k_x - \sigma_y k_y)
    + (M_\nu-B_\nu k^2)\sigma_z,
\end{align}
with the Pauli matrices ($\sigma_x$, $\sigma_y$, and $\sigma_z$) acting on the $(H_1, E_1)$ subspace of each Kramers block. Similarly, the tunnel couplings $V_{\mu}$ for $\mu = \{LC, CR\}$, referring to the left-central and central-right QW couplings, are $V_\mu = v_\mu(\bm{k}) \oplus v_\mu^*(-\bm{k})$, with
\begin{align}
    v_\mu(\bm{k}) &= \dfrac{1}{2}\Big[
        \Delta_{0,\mu} + \Delta_{z,\mu} \sigma_z + \alpha_\mu (\sigma_x k_x - \sigma_y k_y)
        \Big].
    \label{eq:Vcoupling}
\end{align}
All coefficients above are calculated following the $\bm{k}\cdot \bm{p}$ perturbation theory up to second order. These are shown in the Supplementary Material\cite{SupplementalMaterial}.

\section*{Theoretical Results}

In this section discuss the energy dispersions of the TQWs. The different topological phases of the TQW are presented in terms of a phase diagram as a function of the geometric parameters $d_0$ and $t$ [see Fig.~\ref{fig:1}(a)], and labeled by the corresponding spin Chern number. The edge state dispersions are illustrated for representative cases of each phase.

\subsection*{Energy subbands}

Figures \ref{fig:1}(b)-(e) illustrate the band structure for triple HgTe QWs with $t=3$~nm and increasing $d_0 = \{5,6,7,8.5\}$~nm. In Fig.~\ref{fig:1}(b) the system is in the trivial regime i.e., all three conduction subbands have a predominantly $\Gamma_6$ electron-like (E-like) character, while the three valence subbands have a predominantly $\Gamma_8$ hole-like (H-like) character. Here, we have, for each Kramers pair, three non-degenerate conduction E-like subbands ($\left |E_{1\pm}^{-}\right \rangle$, $\left |E_{1\pm}^{0}\right \rangle$, and $\left |E_{1\pm}^{+}\right \rangle$), which arise from the hybridization of the lowest conduction subbands of the left (L), central (C) and right (R) wells, namely, $\left |E_{1\pm}^{L}\right \rangle$, $\left |E_{1\pm}^{C}\right \rangle$ and $\left |E_{1\pm}^{R}\right \rangle$. On the other hand, the three H-like bands (for each Kramers pair) are nearly degenerate because the heavy-hole states $\left |H_{1\pm}^{\nu}\right \rangle$ are strongly localized within each well due to their larger effective mass, and only show significant hybridization for $t < 1$~nm. As we increase $d_0$ in Figures \ref{fig:1}(b)-(e), the E-like subbands  cross the H-like subbands, one by one. Each time a E-like subband crosses down the three H-like subbands (with same Kramer pair index), one H-like band flips the sign of its effective mass, but it still remains nearly degenerate with the other H-like bands at $\bm{k}=0$ due to the small overlap of their wavefunctions. Similarly to the case of only one well, the subband inversions produce topological phase transitions. Here, however, only some of these crossings characterize the phase transitions, and this will be discussed with more detail in the next section. For now, we should only note that the band structures in Figs.~\ref{fig:1}(c)-(d) are topologically non-trivial, but gapless. Moreover, it is only when all the E-like subbands are above [Fig.~\ref{fig:1}(b)] or below [Fig.~\ref{fig:1}(e)] all the H subbands that the system shows a well-defined gap, and as a consequence can be claimed to be either a trivial or topological insulator.

\begin{figure}[ht]
 \centering
 \includegraphics[width=\columnwidth]{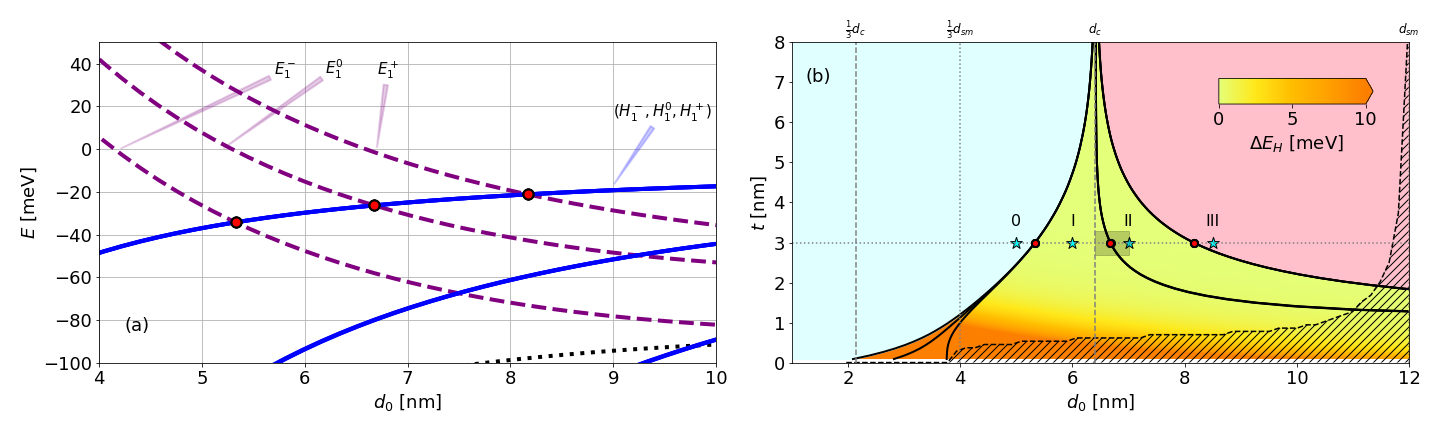}
 \caption{
 (a) Crossings of the E-like $\bm{k}=0$ band edges with the H-like bands as a function of $d_0$, and $t=3$~nm. (b) Phase diagram as a function of $d_0$ and $t$. The black solid lines mark the $E$--$H$ band crossings, and the red circles along the $t=3$~nm line correspond to those in (a). Shaded region marks the SM phase, and the colors within phases I and II refer to hybridization gap between H bands, $\Delta E_H$, which becomes clear only for $t < 1$~nm, and also splits the first solid black line for the $E_1^-$--$H_1$ crossings. The cyan stars along the $t=3$~nm line mark the points (0, I, II, III) corresponding to the phases illustrated in Fig.~\ref{fig:1}(b)-(e). The shaded rectangle near $d_0 = 6.7$~nm and $t=3$~nm marks the parameters of the experimental sample, with the area shaped to illustrate the experimental uncertainty of $\pm 0.3$~nm in both $d_0$ and $t$.
 }
 \label{fig:2}
\end{figure}

It is also important to stress that the valence bands obtained here also show the ``\textit{camel back}'' profile~\cite{doi:10.1126/sciadv.aba4625,Krishtopenko2018Realistic}, thus also presenting an indirect gap defined by $\Delta E_I$ [See Figs.~\ref{fig:1}(b)--(e)]. This feature appears due to the strong hybridization of the QW subbands for large thickness $d_0\gtrsim 7$~nm {(or small $t < 1$~nm)}, where the subbands are close (in energy) to each other [See Fig.~\ref{fig:2}(a)]. Furthermore, for large $d_0 \sim 12$~nm (not shown), the indirect gap $\Delta E_I$ [see Fig.~\ref{fig:1}(b)] closes and the system becomes semi-metallic (SM). Notice that in Fig.~\ref{fig:1}(e) the indirect gap is already smaller than the direct gap. We emphasize that throughout this work, the notation ``\emph{semi-metalic (SM)}'' will refer to systems in which the indirect gap $\Delta E_I$ vanishes, while the ``\emph{gapless}'' will refer to band structures with vanishing gap at $\bm{k}=0$.

\subsection*{Topological phase transition and Chern number}

In principle, within our system we have three conduction subbands crossing three different valence subbands, yielding a total of nine different inversions. However, only three out of those nine inversions give rise to a topological phase transition, and therefore, a precise characterization of the inversions becomes important. For instance, a counter intuitive scenario occurs in InAs/GaSb type-II QWs\cite{PhysRevLett.100.236601}, where the topological phase transition takes place as E-like states localized at the InAs layer crosses H-like states from the GaSb layer. 

It is important to stress that a band inversion is a necessary, but not sufficient ingredient to have a topological phase transition. Accordingly, bands must not only invert, but also hybridize to open a gap between them after the inversion. It turns out that only three out of the nine inversions in the TQW satisfy these conditions. To identify which are the relevant crossings, we diagonalize the effective Hamiltonian from Eq.~(\ref{eq:H2D}) at $\bm{k}_{\parallel}=0$. Using $Q = \{E, H\}$ to label the E-like and H-like subbands, the diagonal subbands for the case of three identical QWs read as
\begin{align}
    \left | Q_{1\pm}^{-1}\right \rangle &= \frac{1}{2}\left | Q_{1\pm}^{L}\right \rangle + \frac{1}{\sqrt{2}}\left | Q_{1\pm}^{C}\right \rangle + \frac{1}{2}\left | Q_{1\pm}^{R}\right \rangle, \\
    \left | Q_{1\pm}^{0}\right \rangle &= \frac{1}{\sqrt{2}}\left | Q_{1\pm}^{L}\right \rangle - \frac{1}{\sqrt{2}} \left | Q_{1\pm}^{R}\right \rangle,\\
    \left | Q_{1\pm}^{+1}\right \rangle &=  \frac{1}{2}\left | Q_{1\pm}^{L}\right \rangle - \frac{1}{\sqrt{2}}\left | Q_{1\pm}^{C}\right \rangle + \frac{1}{2}\left | Q_{1\pm}^{R}\right \rangle.
\end{align}
Projecting the Hamiltonian from Eq.~(\ref{eq:H2D}) into this basis yields
\begin{align}
    \label{eq:H2D-diag}
    \tilde{H}_{2D} &=
    \begin{pmatrix}
     \tilde{H}_{-1} & 0 & 0 \\
     0 & \tilde{H}_{0} & 0 \\
     0 & 0 & \tilde{H}_1
    \end{pmatrix}. 
\end{align}
Here, the block diagonal terms $\tilde{H}_\mu$ with $\mu=\{-1,0,1\}$ have the usual BHZ form with $\tilde{H}_\mu = \tilde{h}_\mu(\bm{k}) \oplus \tilde{h}_\mu^*(-\bm{k})$, and
    \begin{align}
    \tilde{h}_\mu(\bm{k}) =& \; 
    (\tilde{C}_\mu-D k^2) + \tilde{A}_\mu(\sigma_x k_x - \sigma_y k_y)
    + (\tilde{M}_\mu-B k^2)\sigma_z,
\end{align}
with renormalized parameters $\tilde{A}_\mu=A+\mu\alpha/\sqrt{2}$, ${\tilde{C}_\mu=C + \mu\Delta_0/\sqrt{2}}$ and ${\tilde{M}_\mu=M + \mu \Delta_z/\sqrt{2}}$. It is evident from Eq.~(\ref{eq:H2D-diag}) that the only hybridization happens between pairs $\{ \left | E_{1\pm}^{-1}\right \rangle,\left | H_{1\pm}^{-1}\right \rangle \}$, $\{ \left | E_{0\pm}^{1}\right \rangle,\left | H_{0\pm}^{-1}\right \rangle \}$ and $\{ \left | E_{1\pm}^{+1}\right \rangle,\left | H_{1\pm}^{+1}\right \rangle \}$. Accordingly, it follows from the BHZ model \cite{Bernevig2006BHZ, Bernevig2006QSHE} that the topological phase transition will only take place when the energies of these individual pairs invert, \text{i.e.} as each $\tilde{M}_\mu$ changes from positive to negative, the corresponding spin Chern number goes from 0 (trivial) to 1 (topological).

\begin{figure*}[ht]
 \centering
 \includegraphics[width=\textwidth]{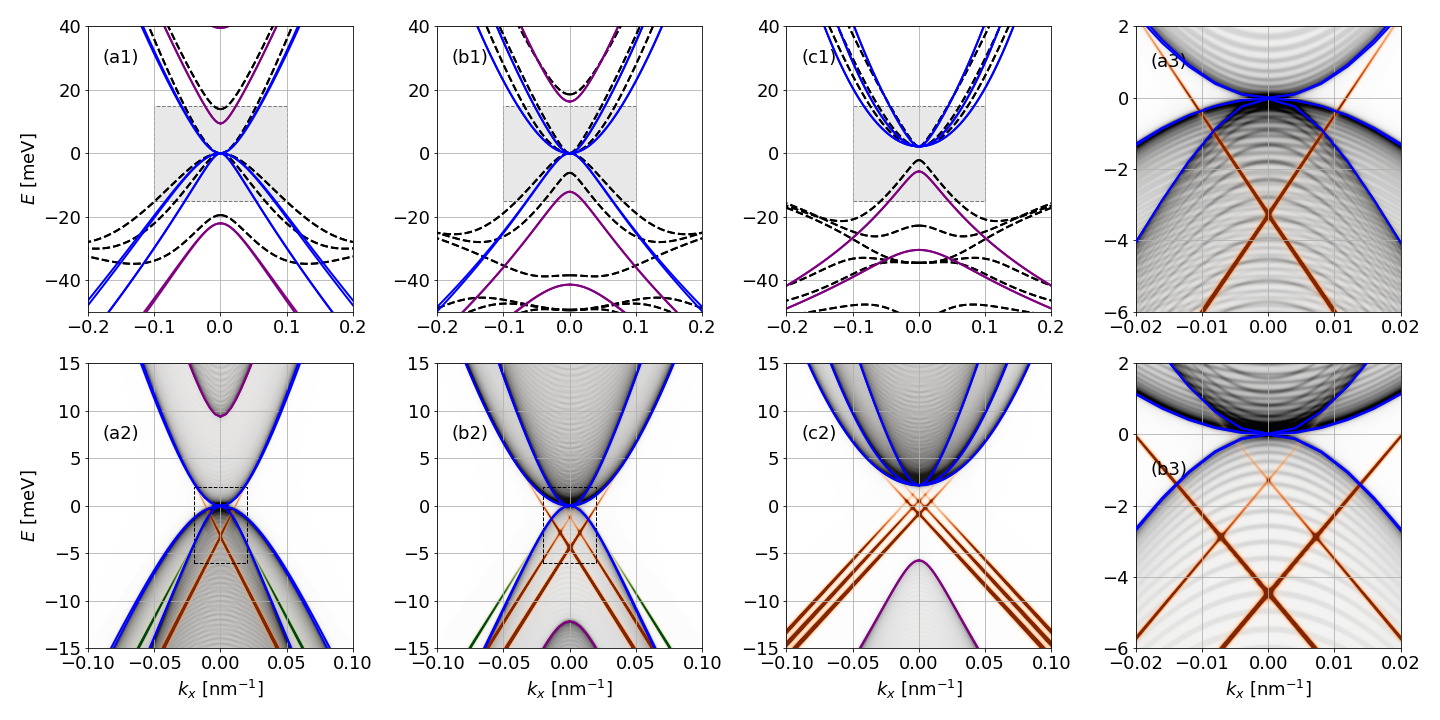}
 \caption{
 (a1, b1, c1) Comparison of the bulk band structures calculated with the $8\times 8$ Kane model (black dashed lines) and the effective 2D \tqwbhz model (solid lines) with E-like bands in purple and H-like bands in blue. The panels correspond to the phases I, II and III indicated by the cyan stars in Fig.~\ref{fig:2}(b). (a2, b2, c2) Band structures for a nanoribbon geometry of width $L_y = 1000$~nm shown by the local density of states of the bulk (gray tones) and edges (orange and green tones). The $E$-$k$ range match the shaded regions from the panels above. The \tqwbhz bulk bands are shown as the solid lines as a guide to the eyes. (a2) For $d_0=6$~nm the toplogical regime with one E-band below the H bands shows one pair of topological edge state (orange) and two pairs of trivial edge states (green). (b2) As a second E-band crosses the H bands for $d_0=7$~nm one gets two topological edge states a single trivial edge state. (c2) For $d=8.5$~nm the system reaches the full topological regime with three topological edge states and a well defined gap. (a3, b3) Zoom into the rectangles of (a2, b2) emphasizing the edge states.
 }
 \label{fig:3}
\end{figure*}

To understand and characterize how the QW hybridizations lead to phase transitions, in Fig.~\ref{fig:2} we draw the corresponding phase diagram as a function of the well width $d_0$ and barrier thickness $t$. First, for fixed $t=3$~nm in Fig.~\ref{fig:2}(a), we see that as the well width increases, the E-like (H-like) subbands move down (up) in energy, a feature well known for the single HgTe/CdTe quantum wells \cite{Bernevig2006QSHE, Bernevig2006BHZ}. The crossings between the E-like and H-like subbands pairs, $\{ \left | E_{1\pm}^{\mu}\right \rangle,\left | H_{1\pm}^{\mu}\right \rangle \}$, are highlighted (red circles) in Fig.~\ref{fig:2}(a) and also in Fig.~\ref{fig:2}(b), along the $t=3$~nm line. In Fig.~\ref{fig:2}(b), the solid black lines mark the parameter values that yield crossings between subband pairs $\{ \left | E_{1\pm}^{\mu}\right \rangle,\left | H_{1\pm}^{\mu}\right \rangle \}$. These lines delimit different regions of our phase diagram, which are labeled by the their corresponding spin Chern number, \textit{i.e.}, 0, I, II and III. This can be clearly seen in Figs.~\ref{fig:1}(b)-(e), which contain the bandstructures corresponding to the cyan stars in Fig.~\ref{fig:2}(b). As a consequence, the region 0 corresponds to the trivial insulator regime, while regions I, II, and III correspond to the topological insulator regimes with one, two and three pairs of topological helical edge states, respectively, despite the gapless character of the full spectrum of both phases I and II for $t\gtrsim 1$~nm. Finally, the shaded region marks the semi-metallic phase, representing the cases where the indirect band gap $\Delta E_I$ closes. The color map represents the nominal value of the gap at $\bm{k}_{\parallel}=0$, which only becomes significant for $t < 1$~nm. 

\subsection*{Edge states}

To illustrate the characteristics of each phase presented above and in Fig.~\ref{fig:2}(b), we now plot and analyze the energy spectrum of representative cases in the presence of an extra confinement along the $y$ direction. To calculate the spectrum the confined system, we consider the \emph{effective} \tqwbhz 2D model presented above, which describes the effective Hamiltonian for the three lowest (highest) conduction (valence) subbands. Additionally, here we consider a hard-wall confinement at $|y| = L_y/2$, with $L_y \sim 1000$~nm. The spectrum is then calculated using a recursive Green's function method \cite{Odashima2016RGF}, which allow us to calculate the local spectral function, $A(E, k_x)$, for the bulk and edges states with an efficient exponential decimation. In Figs.~\ref{fig:3}(a1), (b1) and (c1) we plot $A(E, k_x)$ for the same parameters of Fig.~\ref{fig:1}(c)-(e). We see that for $|k_x|\lesssim 0.1$~nm$^{-1}$, the bands of the effective \tqwbhz 2D model (solid lines) are in good agreement with the bands obtained numerically from the $8\times8$ Kane model (dashed lines), although they are not able to reproduce accurately the ``\textit{camel back}'' profile around $|\bm{k}| \gtrsim 0.2$~nm$^{-1}$. In Figs.~\ref{fig:3} (a2), (b2), (c2) we have the spectrum corresponding to the cyan stars, I, II and III in Fig.~\ref{fig:2}(b), with gray bands representing bulk bands, and orange and green bands representing states localized at edges of our system. Interestingly, it is possible to identify here two different types of edge states. The ones indicated by the orange color are topological edge states with corresponding bands connecting conduction to valence band. Conversely, the ones indicated by the green color are trivial edge states predicted previously in Ref.~[\citen{PhysRevB.98.161111}]. While the topological ones arise from the non-trivial topology of our system, the trivial ones appear when we confine bands that have a strong linear dispersion~\cite{PhysRevB.98.161111}. For this reason, these edge states appear due to the approximately chiral symmetry of the bands~\cite{PhysRevB.98.161111}. Even though the trivial edge states are not protected against backscattering, it was shown that it is possible to make these edge states protected when the ribbon is reduced to a quantum dot~\cite{PhysRevLett.121.256804}.

\section*{Experimental results}
\label{sec:exp}

\begin{figure*}[b!]
 \centering
 \includegraphics[width=\textwidth]{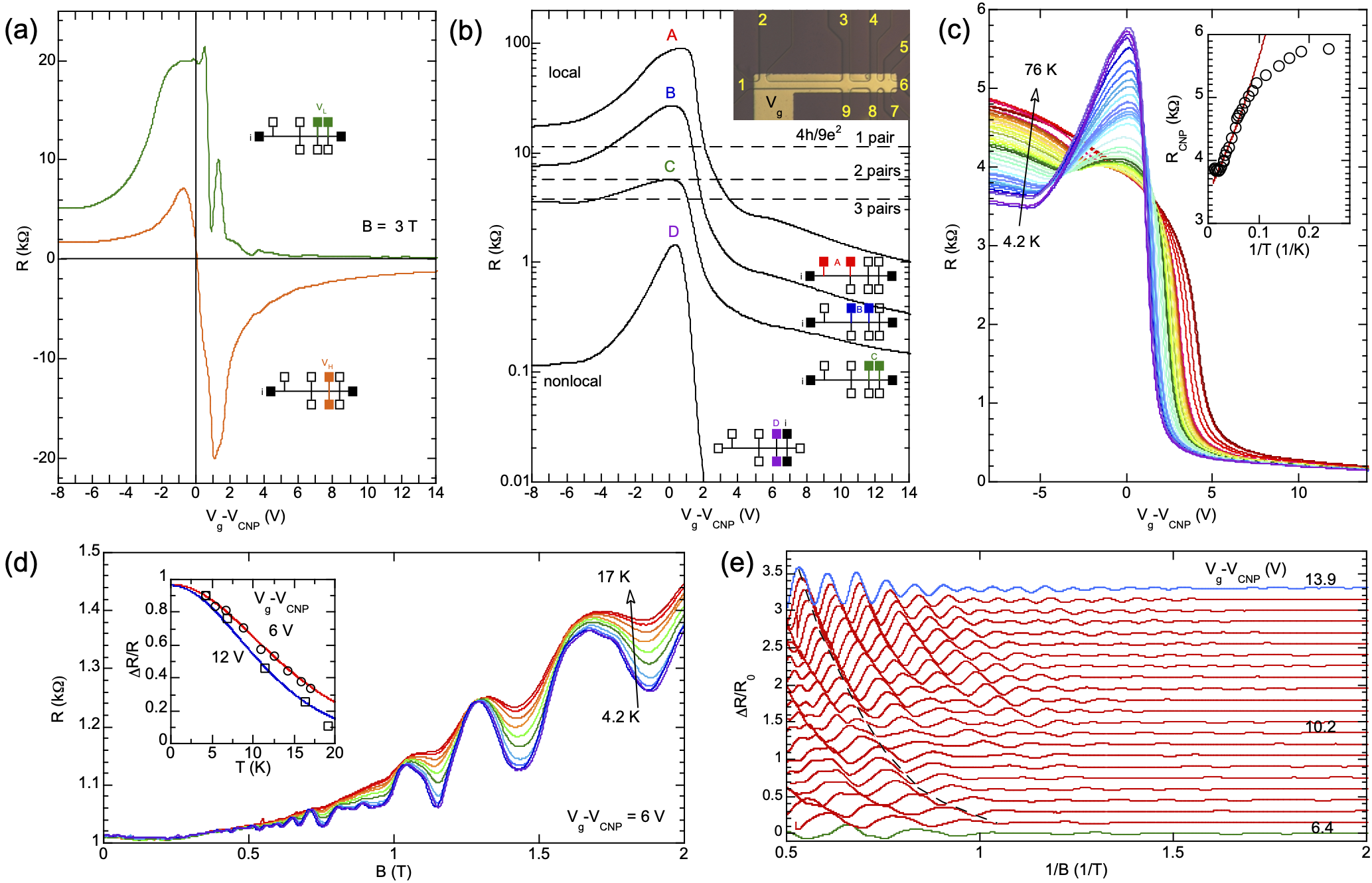}
 \caption{
(a) Longitudinal (green) and Hall (orange) resistances at B = 3 T and T = 4.2 K as a function of gate bias. (b) Local resistance as a function of the gate voltage measured along segments with different lengths (A to C) and nonlocal result (D). Insets show device scheme and measurement configurations. The dashed lines are the expected resistances calculated using Landauer-Buttiker formalism for a device with 9 terminals including the contribution of several pairs of edge states. (c) Resistance as a function of the gate voltage for different temperatures. The inset shows the resistance at CNP as a function of 1/T with solid line corresponding to $R \sim \exp(\Delta/2k_BT )$ with an activation energy\cite{Podgornykh2019Activation} $\Delta = 0.8$~meV. Shubnikov-de Haas oscillations as function of (d) temperature and (e) gate voltage. Inset in (d) shows the fitting of the amplitude dependence at a fixed magnetic field of 1.2 T for two different values of $V_g-V_{CNP}$ where the solid line is a fitting using Lifshitz–Kosevich formula. Curves in (e) were vertically shifted for clarity and the dashed line is a guide to the eye for the change in the oscillations phase.}
 \label{fig:4}
\end{figure*}

Triple quantum wells based on HgTe/Cd$_x$Hg$_{1-x}$Te with [013] surface orientation and equal well widths of $d_0 = 6.7$~nm and barrier thickness $t=3$~nm were prepared by molecular beam epitaxy (MBE). The sample structure is shown in Fig. \ref{fig:1}(a). The layer thickness was determined by ellipsometry during MBE growth, with accuracy of $\pm 0.3$~nm. The devices are multiterminal bars containing three 3.2 $\mu$m wide consecutive segments of different length (2, 8, and 32 $\mu$m) and nine contacts [see inset in Fig. \ref{fig:4}(b)]. The contacts were formed by the burning of indium to the surface of the lithographically defined contact pads. The growth temperature was near 180$^{\circ}$C, therefore, the temperature during contacts fabrication process was relatively low. On each contact pad, the indium diffuses vertically down providing an ohmic contact to all three quantum wells, with contact resistance in the range of 10–50 k$\Omega$. During AC measurements we continuously checked that the reactive component of the impedance never exceeds $5\%$ of the impedance, which demonstrates the good ohmicity of the contacts. The I-V characteristics are also ohmic for low voltages. A dielectric layer, 200 nm of SiO$_2$, was deposited on the sample surface and then covered by a TiAu gate. The density variation with the gate voltage is $\sim 0.9\times 10^{11}$~cm$^{-2}$/V, estimated from the dielectric thickness and in comparison with the calculated frequencies of the SdH oscillations shown in the next section. The transport measurements were performed in the range of temperatures (T) from 1.4 to 80 K by using a standard four-point circuit with 1–13 Hz AC current of 1–10 nA through the sample, which is sufficiently low to avoid overheating effects. 

Three devices from the same substrate were studied. Figure \ref{fig:4}(a) shows the measured transport under strong magnetic field that identifies the charge neutrality point (CNP), with near zero carrier density, in the energy spectrum.  Sweeping the gate voltage (V$_{g}$) from positive to negative values depopulates the electron states and populates the hole states, while the Fermi level passes through the CNP. The longitudinal resistance exhibits oscillating behaviour on the electronic side, however at the hole side the resistance shows monotonic behaviour due to strong scattering between the cone and the heavy hole branches \cite{Kozlov}. In the CNP, the electron-like Hall resistance jumps from the negative quantized value $\sim$ h/e$^{2}$ to the hole-like positive value $\sim$ h/4e$^{2}$. The quantum Hall effect in HgTe TQWs is beyond the scope of this work and will be reported in a forthcoming publication.

Fig. \ref{fig:4}(b) shows local and nonlocal resistance as function of gate voltage in a representative device. In the local case, the current flows between contacts 6-1 and the voltage was measured in the different device segments: 3-2 (curve A), 4-3 (curve B), 5-4 (curve C). For the nonlocal case, the current flows between 7-5 and the voltage measured in 8-4. The resistance maximum for all curves occurs at the CNP, as identified in (a). Curve C, situation closer to the ballistic transport, presents a maximum in agreement with the Landauer-Büttiker calculation for two pairs of edge states in a device with nine terminals. Nevertheless, in the next section, we will discuss that this direct association can be misleading if the bulk contributions are not considered.

In order to identify the nature of the transport in the triple QW sample, we have measured
the temperature dependence of the resistance near the CNP. The variation of the resistance with the gate voltage and temperature is shown in Fig.\ref{fig:4}(c) where the evolution resembles that for single well 2D TIs \cite{PhysRevB.96.045304}. The resistance decreases sharply at T $>$ 15 K while saturating below 10 K, indicating a small mobility gap of 0.8 meV (see inset).

Fig.\ref{fig:4}(d) and (e) shows Shubnikov-de Haas (SdH) oscillations measured in the region of electron conductivity as function of temperature and gate bias, respectively. The inset in Fig.\ref{fig:4}(d) displays that the temperature dependence of the SdH oscillations is well described by the Lifshitz–Kosevich formula. Surprisingly, Fig.\ref{fig:4}(e) presents a strong change in the phase of the SdH oscillations, as indicated by a dashed line that follows a constant phase. Sudden changes in the phase at 8 and 10 V could be related to variations in the Berry phase across system transitions.

Fourier analysis of the magnetoresistance in Fig.\ref{fig:4}(e) is displayed in Fig.\ref{fig:FigNominal}(c). Two peaks corresponding to branches of conduction-band carriers were obtained. The oscillations frequency increased with bias voltage and we observed the splitting of the upper frequency peak for sufficiently high gate voltage. The experimental results are compared with the theoretical model in the following section. 

\section*{Discussion}

The triple well sample used in experiment falls quite close to the phase transition line with $d_0 = (6.7 \pm 0.3)$~nm and $t = (3.0 \pm 0.3)$~nm, as indicated by the shaded rectangle in Fig.~\ref{fig:2}(b). 
These $\pm 0.3$~nm uncertainties in $d_0$ and $t$ are sufficient to locally shift the system between phases I and II along the sample. Nevertheless, here we focus on the nominal geometrical parameters $d_0 = 6.7$~nm and $t=3$~nm, and consider the fluctuations qualitatively in the following discussion.

\begin{figure}[t]
 \centering
 \includegraphics[width=\columnwidth]{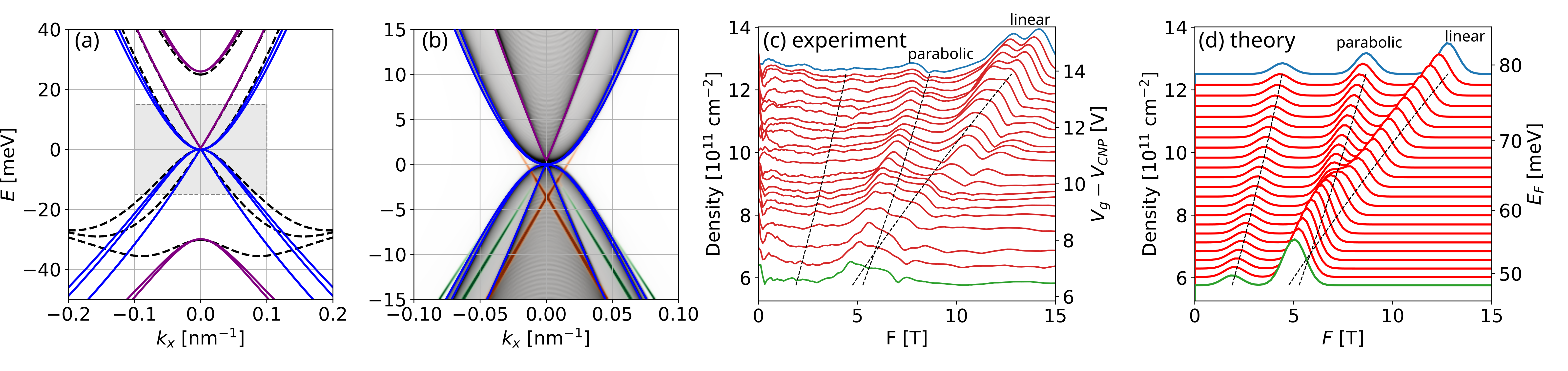}
 \caption{(a) Bulk and (b) edge energy dispersions for the nominal experimental parameters of the triple quantum well, \textit{i.e.,} $d_0 = 6.7$~nm and $t=3.0$~nm, which falls close the the phase transition line, shaded rectangle in Fig.~\ref{fig:2}(b). The lines and colors follow the same definitions as in Fig.~\ref{fig:3}. The system is in phase I, with the $E_1^0$ and $H_1^0$ bands hybridized with nearly vanishing mass $\tilde{M}_0$, yielding a linear pair of Dirac bands. (c) The Fourier transform of the measured Shubnikov–de Haas oscillations show two main peaks splitting with increasing $V_g$ and a further Rashba-like splitting at high $V_g$. (d) The theoretical SdH frequencies $F$ for the linear and parabolic bands from panel (a) qualitatively matches the experimental measurements. In both (c) and (d), the dashed lines mark $F_j(E_F)$ for each band with $E>0$ from panel (a) and the peaks are built from gaussian broadenings for easy comparison with panel (c).}
 \label{fig:FigNominal}
\end{figure}

As shown in Fig.~\ref{fig:FigNominal}, the second E-like subband $E_1^0$ and the H-like $H_1^0$ subband are nearly crossing, \textit{i.e.} $\tilde{M}_0 \approx 0$, forming a Dirac-like dispersion with a small gap of $\tilde{M}_0 \sim 0.6$~meV, which is close to the experimental activation energy of $0.8$~meV. The remaining two $H$ bands form parabolic bands with positive and negative effective masses. For a slightly larger $d_0$ or $t$ (within the $\pm 0.3$~nm experimental uncertainty) the $E_1^0$ and $H_1^0$ bands would cross each other to yield the phase II regime ($\tilde{M}_0 < 0$). Nevertheless, the inverted gap in phase II would still be small ($\tilde{M}_0 \sim -0.6$~meV) and these bands would still have a nearly linear Dirac dispersion. Moreover, with a small negative gap, the corresponding edge states would not be well defined, since its localization length is inversely proportional to this gap. Therefore, within the experimental uncertainty for $d_0$ and $t$, we would expect to have only one pair of well defined edge states near the phase transition line from phase I to II.

\subsection*{Local and non-local resistivities}

To analyze the transport properties of the system in the experimental setup, first, let us consider the ballistic regime and ignore the bulk contributions to the local and non-local transport measurements. From a nine terminals Landauer-Büttiker geometry, one would expect the local and non-local resistances to be
\begin{align}
    R_{i:6,1}^{v:5,4} = R_L &= \dfrac{4}{9n} \dfrac{h}{e^2} \approx \dfrac{11.1}{n} \text{ k}\Omega,
    \\
    R_{i:7,5}^{v:8,4} = R_{NL} &= \dfrac{10}{9n} \dfrac{h}{e^2} \approx \dfrac{27.7}{n} \text{ k}\Omega,
\end{align}
where $n$ is the number of edge states in each edge of the sample, and the labels $i:$ and $v:$ indicate the contacts used to apply the currents and measure the voltages, respectively, which were presented in the previous section. As discussed above, here we expect to have only $n=1$ pairs of well defined edge states. However, as seen in Fig.~\ref{fig:4}(b), both $R_L$ and $R_{NL}$ deviate significantly from this ideal model for $n=1$. Indeed, these deviations are justified by the bulk contributions for transport, since the edge states are immersed in the nearly gapless bulk, as seen in Fig.~\ref{fig:FigNominal}(b). More specifically, this characteristic gives to opposite contributions. First, notice that the $R_L$ measurements in Fig.~\ref{fig:4}(b) decreases as the distance between contacts decreases, which is expected to asymptotically approaches the ballistic regime for short distances. Second, at the shortest distance, line C falls below the expected line for $n=1$ pairs of edge states. We interpret this as a consequence of finite bulk conductivity leading to reduced $R_L$ and $R_{NL}$. In summary, we expect scattering effects to lead to increased resistivities, while the bulk conductivity contributes to reduce the resistivities. These contrasting contributions show that transport measurements are ineffective to characterize if the system is in phase I or II.

\subsection*{Shubnikov–de Haas oscillations}

Shubnikov-de Haas (SdH) oscillations are oscillations of the 2D magnetoresistivity of a material as a function of the external magnetic field $B$~\cite{sdh-original,sdh-original2,lifshitz1956theory,ihn10:book}. They were discovered in early 1930 and currently are one of the most important tools to access and extract values for both the 2D electronic and hole semiconductor densities at $B=0$~\cite{ihn10:book,PhysRevLett.51.126}, the difference between electronic density of different subbands~\cite{PhysRevLett.51.126,PhysRevLett.84.713}, the electron and hole effective masses~\cite{PhysRevLett.51.126,ihn10:book,PhysRevX.7.031010} and also spin-orbit couplings e.g., Rashba~\cite{PhysRevB.41.7685,PhysRevLett.78.1335,PhysRevX.7.031010}. Most recently, SdH oscillations have also been used to probe the 2D Dirac-like character of both graphene energy dispersion  \cite{zhang2005experimental} and 2D Dirac-like surface states of 3D topological insulators \cite{Ren2010,PhysRevB.86.045314}. For low temperatures (\textit{i.e.,} $k_BT \ll \mu \approx E_F$), the SdH oscillations are well described by the Lifshits-Kosevich (LK) \cite{lifshitz1956theory} expression,  which in the absence of Zeeman and spin-orbit couplings reads~\cite{lifshitz1956theory,PhysRevB.44.3793,ihn10:book,Gusynin2005SdH, Raichev2020SdH}
\begin{align}
    \Delta \sigma_{xx}^{(j,\sigma)} &\propto 2 \sum_{\ell=1}^{\infty}
    \dfrac{\ell\lambda_{j,\sigma}}{\sinh(\ell\lambda_{j,\sigma})}
    e^{-\ell \tau_{j,\sigma}/\tau_0}
    \cos\left[
    2\pi\ell\left(
    \dfrac{F_{j,\sigma}}{B} + \phi_{j,\sigma}
    \right)
    \right],\label{cos}
    \\
    \lambda_{j,\sigma} &= \dfrac{4\pi^3\hbar k_BT}{eB}g_{j,\sigma}(E_F),
    \label{eq:lambda}
    \\
    \tau_{j,\sigma} &= \dfrac{2\pi^2 \hbar^2}{eB}g_{j,\sigma}(E_F),
    \\
    F_{j,\sigma} &= \dfrac{2\pi\hbar}{e}n_{2D,{j,\sigma}}(E_F), \label{Fs}
\end{align}
where $\Delta \sigma_{xx}^{(j,\sigma)} = \sigma_{xx}^{(j,\sigma)}(B)-\sigma_{xx}^{(j,\sigma)}(0)$ is the differential longitudinal conductivity for spin $\sigma = \{\uparrow, \downarrow\}$ and band index $j$. For the experiments analyzed in this work, the most relevant bands are the first two nearly spin-degenerate conduction bands shown in Fig.~\ref{fig:FigNominal}(a)-(b), i.e., one linear (L) and one parabolic (P) band, and hence we set $j= \{L, P\}$, with corresponding energies $\varepsilon_L = \pm \hbar v_F k$, and $\varepsilon_P = \hbar^2 k^2/2m^*$. In this generic form, the temperature dependent term ($\lambda_{j,\sigma}$) and the Dingle factor ($\tau_{j,\sigma}$) are written in terms of the density of states of each band at the Fermi level $E_F$, $g_{j,\sigma}(E_F)$. The frequency $F_{j,\sigma}$, determining the SdH oscillations with respect to $1/B$ is written in terms of the charge density $n_{2D,{j,\sigma}}(E_F) \approx \int_0^{E_F} g_{j,\sigma}(\varepsilon)d\varepsilon$. The characteristics of the energy dispersion of each band enters within $g_{j,\sigma}(E_F)$ and $n_{2D,{j,\sigma}}(E_F)$. Accordingly, for the linear band we have $g_{L,\sigma}(E_F) = E_F/(2\pi \hbar^2 v_F^2)$ and ${n_{2D,L,\sigma}(E_F) = E_F^2/(4\pi\hbar^2v_F^2)}$, while for the parabolic band we have $g_{P,\sigma}(E_F) = m^*/(2\pi\hbar^2)$, and $n_{2D,P,\sigma}(E_F) = m^*E_F/(2\pi\hbar^2)$. However, the main difference between the linear (Dirac-like) and the parabolic bands lies in the phase $\phi_j$, which assumes the value $\phi_P = 1/2$ for parabolic bands and $\phi_L = 0$ for linear bands. For the case of linear bands, the absence of any phase within the argument of $\cos(...)$ in Eq.~(\ref{cos}) is interpreted due to the non-zero ($\pi$) Berry's phase of 2D Dirac cones~\cite{PhysRevLett.82.2147,PhysRevB.69.075104,PhysRevB.71.125124,PhysRevB.84.035301}. This introduces an extra $\pi$ phase within the corresponding $\cos(...)$ argument, which effectively cancels out when combined with the previous existed $\pi$ phase. Overall, the generic form of $F_j \propto n_{2D,j}$ tell us that the experimentally measured frequencies can be used to obtain the different electronic densities of our bands, independently of its linear or parabolic dispersion. This allows us to directly compare the experimental data given by Fig.~\ref{fig:FigNominal}(c) to a theoretical model in terms of a common total density axis, which is used to shift the lines for each SdH curve in Figs.~\ref{fig:FigNominal}(c) and (d). 

In practice, the experimental measurements of the SdH oscilattions are done through the total differential resistivity $\Delta R/R_0 \propto \rho_{xx}$. From the experimental data within Fig.~\ref{fig:4}(a), we see that, for $V_g - V_{CNP}\gtrsim 4$~V, we have $\rho_{xx}\gg \rho_{xy}$. This translates to $\sigma_{xy}\ll \sigma_{xx}$, which yields $\sigma_{xx} \approx -\rho_{xx}/\rho_{xy}^2$. A direct consequence of this approximation\cite{ Ando_2013} is that, now, $\Delta R/R_0$ can be written as a weighted sum of the LK expressions above over our different bands with corresponding different amplitudes $P_{j,\sigma}$, i.e.,
\begin{align}
    \dfrac{\Delta R}{R_0} &\approx
    \sum_{j,\sigma} P_{j,\sigma} \Delta \sigma_{xx}^{(j,\sigma)},
    \label{Rtot} \\
    P_{j,\sigma} &= \dfrac{\sigma_{xx}^{(j,\sigma)}(0)}{\sum_{j',\sigma'}\sigma_{xx}^{(j',\sigma')}(0)} \approx \dfrac{g_{j,\sigma}}{\sum_{j',\sigma'}g_{j',\sigma'}}. \label{amplitude}
\end{align}
Once we have now obtained the theoretical equations that are going to be compared to the experimental quantities, we move to the experimental data.

In Fig.~\ref{fig:FigNominal}(c) we plot the Fourier transformation of the SdH oscillations of Fig.~\ref{fig:4}(e) for different values of $V_g - V_{CNP}$. The peaks of Fig.~\ref{fig:FigNominal}(c) correspond to the different SdH frequencies $F_{j,\sigma}$ [Eq.~(\ref{Fs})]. Although the SdH frequencies are experimentally obtained as a function of different $V_g$, which controls the Fermi energy, here we introduce a voltage-to-density conversion factor of $\sim 0.9\times 10^{11}$cm$^{-2}$/V. This is required in order to perform a comparison between our theoretical results to the experimental data. Moreover, this conversion factor shows to be in agreement with previous publications in similar devices \cite{Gusev2020DW,Gusev2021DW}. Accordingly, we also plot the frequencies for different total density $n_{2D} = \sum_j n_{2D,j}$ (left-hand side axis).

In order to compare the experimental results with our theory, in Fig.~\ref{fig:FigNominal}(d) we plot the SdH frequencies ($F_{j,\sigma}$) predicted by our calculations as a function of $E_F$ (or the total 2D electronic density $n_{2D}$). They are obtained using the parameters extracted from the conduction bands dispersions within Fig.~\ref{fig:FigNominal}(a). Namely, we obtain $m^* = 0.0127m_0$ for the parabolic H-like conduction band, and $\hbar v_F = 400$~meV.nm for the E-like linear conduction band. For small $E_F$, we see that both frequencies $F_{L,\sigma}$ and $F_{P,\sigma}$ are close to each other. However, as $E_F$ is increased there is an evident separation between them, stemming from their different $E_F$-dependencies, i.e., $F_{L,\sigma} \propto n_{2D,L,\sigma} \propto E_{F}^2$ and $F_{P,\sigma} \propto n_{2D,P,\sigma} \propto E_F$. Surprisingly, our theoretical frequencies $F_{j,\sigma}$, plotted within Fig.~\ref{fig:FigNominal}(d), present a good agreement with the experimental data shown in Fig.~\ref{fig:FigNominal}(c). This shows that the SdH obtained in the experiments are consistent with the coexistence of a linear and parabolic band, similarly to the case of trilayer graphene~\cite{doi:10.1126/sciadv.aax6550}. This coexistence can also be evidenced by the different dependence of $F_{j,\sigma}$ with respect to $E_F$, which is shown by the black dashed lines in both Figs.~\ref{fig:FigNominal} (c) and (d) as a guide to our eyes. We emphasize that the only adjustable parameter between the experiment and theory presented in Figs.~\ref{fig:FigNominal}(c) and (d) is the voltage-to-density conversion factor ($\sim 0.9\times 10^{11}$cm$^{-2}$/V). Additionally, there are other features of the experimental data that corroborates the evidence of linear and parabolic bands contributing to the SdH oscillations. We discuss them in the next paragraphs.

First, there is an apparent crossing of the two frequency peaks near $F = 6$~T within Fig.~\ref{fig:FigNominal}(c). Since $F_{j,\sigma} \propto n_{2D,j,\sigma}$, such a crossing can only happen if there is an equivalent crossover in the densities $n_{2D,j,\sigma}$ of each band. For a pair of parabolic bands, this would require one of the bands to be at a higher energy, and with a heavier mass (larger DOS). This is unlikely for our sample. In contrast, this crossing of frequencies is expected from the coexistence of linear and parabolic bands nearly degenerated at $\bm{k} = 0$. From the condition $F_{P,\sigma} = F_{L,\sigma}$, it follows that the crossing point occurs for $E_F = 2 m^* v_F^2$, where $m^*$ is the effective mass of the parabolic band, and $v_F$ is the Fermi velocity of the linear band. 

Secondly, accordingly to Eqs.~(\ref{Rtot}) and (\ref{amplitude}) the different amplitudes $P_{j,\sigma}$ affect the relative spectral weight of the Fourier frequencies ($F_{j,\sigma}$). More specifically, for the parabolic band, the DOS, $g_{P,\sigma}(E_F)$, is constant as a function of $E_F$, while for the linear band we have $g_{L,\sigma}(E_F) \propto E_F$. As a consequence, as we increase $E_F$ (or $n_{2D}$), this leads to an increase of the differential resistivity of the linear band, thus increasing its spectral weight. Surprisingly, this is also seen within the experimental data of Fig.~\ref{fig:FigNominal}(c). While the spectral weight of $F_{P,\sigma}$ remains nearly constant for different $E_F$, the spectral weight of $F_{L,\sigma}$ increases as a function of $E_F$.

Thirdly, the temperature dependence of SdH oscillations amplitudes show an indirect evidence of the coexistence of the linear and parabolic bands. From Eq.~\eqref{cos}, we see that the temperature dependence is set by the $\lambda_{j,\sigma}$ terms of each band. However, one cannot experimentally distinguish the individual contribution from each band. Instead, the measurements shown in the inset of Fig.~\ref{fig:4}(d) are adjusted to be fitted by an effective LK expression with a single effective $\bar{\lambda}$, defined by the replacement $g_{j,\sigma}(E_F) \rightarrow \bar{g}(E_F)$. Here, we can expect that this effective DOS $\bar{g}(E_F)$ should lie within a range set by the theoretical $g_{P,\sigma}$ and $g_{L,\sigma}$. Indeed, for $V_g - V_{CNP} = 6$~V, we obtain from the experimental data $\bar{g} = 0.07 \times 10^{11}$~meV$^{-1}$cm$^{-2}$, while the theoretical expression, at the corresponding $E_F$, gives us $g_{P,\sigma} = 0.05 \times 10^{11}$~meV$^{-1}$cm$^{-2}$ and $g_{L,\sigma} = 0.09 \times 10^{11}$~meV$^{-1}$cm$^{-2}$. Moreover, for $V_g - V_{CNP} = 12$~V the experimental measurements result in a slightly increased $\bar{g} = 0.09 \times 10^{11}$~meV$^{-1}$cm$^{-2}$. Since $g_{P,\sigma}$ for the parabolic band is constant as a function of $E_F$, the increase in $\bar{g}$ comes from the linear band, which is theoretically calculated and yields $g_{L,\sigma} = 0.14 \times 10^{11}$~meV$^{-1}$cm$^{-2}$ for  $V_g - V_{CNP} = 12$~V. In both $V_g - V_{CNP} = 6$ and $12$~V cases, the effective $\bar{g}$ lies within the expected range $g_{P,\sigma} < \bar{g} < g_{L,\sigma}$, and it increases with $E_F$ due to the linear band contribution.

To finish a complete characterization of the SdH oscillations, we would need to analyze the phases $\phi_{j,\sigma}$ in the oscillations of Fig.~\ref{fig:4}, and also the beating patterns that typically arise from spin-orbit couplings (SOC). However, current models for the SdH oscillations on linear Dirac bands are limited. Typically, the models neglect the Zeeman splitting (for being small), and do not include the spin-orbit coupling \cite{PhysRevB.69.075104,PhysRevB.71.125124,PhysRevB.84.035301}. Interestingly, it has been discussed that the Zeeman splitting introduces a phase correction to the SdH oscillations for Dirac bands \cite{Ren2010, Xiong_2012, Ando_2013}, but the effect of spin-orbit coupling is unknown at the moment and it cannot be inferred from knowledge of its counterpart in parabolic bands. Since these developments are beyond the scope of this paper, we will leave a discussion of the phase $\phi$ and the splitting of the peaks in Fig.~\ref{fig:FigNominal}(c) at high densities for future works.

\section*{Conclusions}

In summary, we have investigated the phase diagram of triple HgTe quantum wells as a function of its geometric parameters and compared its prediction with experimental measurements for a sample that falls quite close to a phase transition line. For the theoretical investigation of the phase diagram we have projected the 3D Kane Hamiltonian into an effective \tqwbhz 2D model that allowed us to investigate its edge state characteristics in each topological phase. We found that phases I and II are gapless due to small hybridization of H-like states from different quantum wells, but still present, respectively, one and two pairs of edge states immersed in the bulk. It is only in phase III that all E and H bands are inverted and the three sets of edge states form within a bulk gap.

The experimental data, for a sample with geometric parameters that fall quite close to the transition from phase I to II, allowed us to analyze the predictions of the theoretical model and the consequences of having the edge states immersed in bulk. We have seen that non-local resisitivity measurements show a reduced signal due to bulk conductivity, while the local resistivity deviates from perfect quantization due to both bulk conductivity and non-ballistic transport. Consequently, models for edge states within bulk would have to account for these features to achieve reliable comparison with experiments. More interestingly, we have seen that SdH measurements show signatures of the predicted bulk bands given by a set of linear and parabolic bands near the phase transition. However, future work is needed to properly characterize the SdH phase $\phi$ for linear bands in the presence of strong Zeeman and spin-orbit couplings.

\bibliography{main} 

\section*{Acknowledgments}

The authors acknowledge funding from the Brazilian agencies: the National Council for Scientific and Technological Development (CNPq), the Coordination for the Improvement of Higher Education Personnel (CAPES). G.J.F. also acknowledges support from the Minas Gerais Research Foundation (FAPEMIG). F.G.G.H. and G.M.G. acknowledge support from the São Paulo Research Foundation (FAPESP), under grants No. 2015/16191-5 and No. 2018/06142-5. D.R.C. acknowledges support provided by the Center for Emergent Materials, an NSF MRSEC under Award No. DMR-1420451.

\section*{Author Contributions}
G.J.F. and D.R.C. have developed the models and investigated the characteristics the phase diagram and SdH oscillations. 
F.G.G.H., E.B.O. and G.M.G. performed the experimental measurements and data analysis. 
E.B.O., N.N.M., and S.A.D. have grown the studied sample and prepared the experimental devices. 
G.J.F. wrote the manuscript with inputs from D.R.C., F.G.G.H. and G.M.G.
All authors have contributed to discussions and revised the manuscript.

\section*{Competing interests}
The authors declare no competing interests.

\end{document}


\flushbottom
\maketitle
\thispagestyle{empty}

\section*{\texorpdfstring{$8\times8$ Kane model and material parameters}{8x8 Kane model and material}}
\label{app:kane8x8}

We use the Kane Hamiltonian~\cite{bir1974symmetry, Bastard1988, voon2009kp, winkler2003spin} to obtain the subbands of the QWs described in this work. This Hamiltonian it is written in the basis $\left|\Gamma_{6},\frac{1}{2},\pm\frac{1}{2}\right\rangle$, $\left|\Gamma_{8},\frac{3}{2},\pm\frac{1}{2}\right\rangle$, $\left|\Gamma_{8},\frac{3}{2},\pm\frac{3}{2}\right\rangle$ and $\left|\Gamma_{7},\frac{1}{2},\pm\frac{1}{2}\right\rangle$ and reads as
\begin{equation}
{\mathcal H}^{8\times8}=\left(\begin{array}{cccccccc}
T & 0 & -\frac{1}{\sqrt{2}}Pk_{+} & \sqrt{\frac{2}{3}}P\hat{k}_{z} & \frac{1}{\sqrt{6}}Pk_{-} & 0 & -\frac{1}{\sqrt{3}}P\hat{k}_{z} & -\frac{1}{\sqrt{3}}Pk_{-}\\
0 & T & 0 & -\frac{1}{\sqrt{6}}Pk_{+} & \sqrt{\frac{2}{3}}P\hat{k}_{z} & \frac{1}{\sqrt{2}}Pk_{-} & -\frac{1}{\sqrt{3}}Pk_{+} & \frac{1}{\sqrt{3}}P\hat{k}_{z}\\
-\frac{1}{\sqrt{2}}Pk_{-} & 0 & U+V & -\bar{S}_{-} & R & 0 & \frac{1}{\sqrt{2}}\bar{S}_{-} & -\sqrt{2}R\\
\sqrt{\frac{2}{3}}P\hat{k}_{z} & -\frac{1}{\sqrt{6}}Pk_{-} & -\bar{S}_{-}^{\dagger} & U-V & C & R & \sqrt{2}V & -\sqrt{\frac{3}{2}}\tilde{S}_{-}\\
\frac{1}{\sqrt{6}}Pk_{+} & \sqrt{\frac{2}{3}}P\hat{k}_{z} & R^{\dagger} & C^{\dagger} & U-V & \bar{S}_{+}^{\dagger} & -\sqrt{\frac{3}{2}}\tilde{S}_{+} & -\sqrt{2}V\\
0 & \frac{1}{\sqrt{2}}Pk_{+} & 0 & R^{\dagger} & \bar{S}_{+} & U+V & \sqrt{2}R^{\dagger} & \frac{1}{\sqrt{2}}\bar{S}_{+}\\
-\frac{1}{\sqrt{3}}P\hat{k}_{z} & -\frac{1}{\sqrt{3}}Pk_{-} & \frac{1}{\sqrt{2}}\bar{S}_{-}^{\dagger} & \sqrt{2}V & -\sqrt{\frac{3}{2}}\tilde{S}_{+}^{\dagger} & \sqrt{2}R & U-\Delta & C\\
-\frac{1}{\sqrt{3}}Pk_{+} & \frac{1}{\sqrt{3}}P\hat{k}_{z} & -\sqrt{2}R^{\dagger} & -\sqrt{\frac{3}{2}}\tilde{S}_{-}^{\dagger} & -\sqrt{2}V & \frac{1}{\sqrt{2}}\bar{S}_{+}^{\dagger} & C^{\dagger} & U-\Delta
\end{array}\right),
\label{eq:kane8x8}
\end{equation}
with
\begin{align}
    T &= E_{c} + \dfrac{\hbar^2}{2m_0}
    \left[ (2F+1)k_{\parallel}^{2} + \hat{k}_z (2F+1) \hat{k}_z
    \right] + a_c(2\varepsilon_{xx} + \varepsilon_{zz}),
\\
    U &= E_{v} -\dfrac{\hbar^{2}}{2m_{0}}\left(\gamma_{1}k_{\parallel}^{2}+\hat{k}_z\gamma_{1}\hat{k_z}\right) + a_v(2\varepsilon_{xx} + \varepsilon_{zz}),
\\
    V &= -\dfrac{\hbar^{2}}{2m_{0}}(\gamma_{2}k_{\parallel}^{2}-2\hat{k}_z\gamma_{2}\hat{k}_z) + b(\varepsilon_{xx} - \varepsilon_{zz}),
\\
    R &= -\dfrac{\hbar^{2}}{2m_{0}}\sqrt{3}(\mu k_{+}^{2}-\bar{\gamma}k_{-}^{2}),
\\
    \bar{S}_{\pm} &= -\dfrac{\hbar^{2}}{2m_{0}}\sqrt{3}k_{\pm}\left(\left\{ \gamma_{3},\hat{k}_z\right\} +\left[\kappa,\hat{k}_z\right]\right),
\\
    \tilde{S}_{\pm} &= -\dfrac{\hbar^{2}}{2m_{0}}\sqrt{3}k_{\pm}\left(\left\{ \gamma_{3},\hat{k}_z\right\} -\dfrac{1}{3}\left[\kappa,\hat{k}_z\right]\right),
\\
    C &= \dfrac{\hbar^2}{m_0}k_{-}[\kappa,\hat{k}_z].
\end{align}
Here, all the parameters depend on the coordinate $z$, which is set along the QW growth direction.
$E_{c}$ and $E_{v}$ correspond to the energies at $k_x=k_y=0$ of the $\left|\Gamma_{6}\right\rangle$ and $\left|\Gamma_{8}\right\rangle$ bands, respectively, and $\Delta$ quantifies the split-off energy separation, i.e., the energy between $\left|\Gamma_{8}\right\rangle$ and $\left|\Gamma_{7}\right\rangle$ bands. Here, $\left[A,B\right]=AB-BA$ denotes the commutator between the $A$ and $B$ operators while $\left\{ A,B\right\} =AB+BA$ denotes their anti-commutator. $P$ is the Kane parameter, $m_{0}$ is the free electron mass, $\gamma_{i}$ are the Luttinger parameters\cite{LuttingerKohn}, which together with $\kappa$ and $F$ accounts for the effective mass correction due to the remote bands. 

\begin{table}[hb!]
    \centering
    \caption{Parameters used in the $8\times 8$ Kane Hamiltonian for HgTe/Hg$_{1-x}$Cd$_x$Te heterostructures. For a concentration $x$ the alloy gap is $E_g(x) = (1-x)E_g^{\rm HgTe} + xE_g^{\rm CdTe} - 0.132x(1-x)$, while all other parameters follow a linear interpolation with $x$. Additionally, we have $P^2=\hbar^2 E_p/2m_0$, $E_c=E_g+E_v$, $\mu = (\gamma_3-\gamma_2)/2$, $\bar{\gamma} = (\gamma_3+\gamma_2)/2$, $\varepsilon_{xx} = [a^{\rm CdTe} - a(x)]/a(x)$, and $\varepsilon_{zz} = -2C_{12} \varepsilon_{xx}/C_{11}$.}
    \begin{tabular}{c|c|c||c|c|c}
    \hline
    Parameter & HgTe & CdTe & Parameter & HgTe & CdTe
    \\ \hline\hline
    $E_p$~[eV] & 18.8 & 18.8 & 
    $a$~[\AA] & 6.46 & 6.48
    \\
    $E_g$~[eV] & -0.303 & 1.606 &
    $a_c$~[eV] & -2.380 & -2.925
    \\
    $E_v$~[eV] & 0 & -0.570 &
    $a_v$~[eV] & 1.31 & 0
    \\
    $\Delta$~[eV] & 1.08 & 0.91 &
    $b$~[eV] & -1.5 & -1.2
    \\
    $F$ & 0 & -0.09 &
    $C_{11}$~[GPa] & 53.6 & 53.6
    \\
    $\gamma_1$ & 4.1 & 1.47 &
    $C_{12}$~[GPa] & 36.6 & 37
    \\
    $\gamma_2$ & 0.5 & -0.28 &
     &  & 
    \\
    $\gamma_3$ & 1.3 & 0.03 &
     &  & 
    \\
    $\kappa$ & -0.4 & -1.31 &
     &  & 
    \\ \hline
    \end{tabular}
    \label{tab:parameters}
\end{table}

\section*{Effective 2D model coefficients}
\label{app:2Dcoeffs}

The coefficients for the effective 2D model\cite{Michetti2012DW, Michetti_2013} are calculated from the $\bm{k}\cdot\bm{p}$ method in the basis of \textit{single QW eigenstates} from the $(k_x, k_y)=0$ numerical solutions of the $8\times8$ Kane model, which we express here as
\begin{equation}
\mathcal{H}^{8\times8} = H_0 + H_x k_x + H_y k_y + H_{xx} k_x^2 + H_{yy} k_y^2 + H_{xy} k_x k_y,
\end{equation}
such that the indexes $\mu$ in each $H_\mu$ indicate their corresponding $k_x$ and $k_y$ powers, while their $z$ and $k_z = -i\partial_z$ dependence is implied within each term. This notation is useful to express the matrix elements in the following discussion by explicitly showing the $(k_x, k_y)$ terms, while omitting the $z$ directions that is integrated on each matrix element. 

Following the Löwdin partitioning, we calculate the effective model defining a set $A$ of eigenstates from the E1 and H1 solutions for the QW $\nu$, which we label as $A = \{\ket{H_{1\pm}^\nu}, \ket{E_{1\pm}^\nu}\}$, and a set $B$ of \emph{remote bands} as all remaining eigenstates $\ket{b}$. In practice, the sum over the remote bands $b \in B$ is truncated up to $\sim 20$ eigenstates above and below the Fermi energy, and its convergence is verified. Additionally, we emphasize that the basis sets A and B are extracted from single QW models of $\mathcal{H}^{8\times8}$, while, in the matrix elements below, $\mathcal{H}^{8\times8}$ is set as the triple QW. Therefore, the basis states are only approximately eigenstates of $H_0$. The resulting model was shown in the main text, and up to second order in Löwdin's perturbation, the matrix elements of the effective Hamiltonian for $a \in A$ are
\begin{align}
 C_\nu + M_\nu &\approx \bra{E_{1\pm}^\nu} H_0 \ket{E_{1\pm}^\nu} + \delta C_\nu + \delta M_\nu,
 \\
 C_\nu - M_\nu &\approx  \bra{H^\nu_{1\pm}} H_0 \ket{H^\nu_{1\pm}} + \delta C_\nu - \delta M_\nu,
 \\
 A_\nu &\approx \bra{E^\nu_{1\pm}} H_x \ket{H^\nu_{1\pm}},
 \\
 D_\nu + B_\nu &\approx \bra{E^\nu_{1\pm}} H_{xx} \ket{E^\nu_{1\pm}}
 + \sum_{b \in B} 
 \dfrac{
    \bra{E_{1\pm}^\nu} H_x \op{b} H_x \ket{E_{1\pm}^\nu}
 }{\varepsilon^0_{E_{1\pm}^\nu}-\varepsilon^0_b},
\\
 D_\nu - B_\nu &\approx \bra{H^\nu_{1\pm}} H_{xx} \ket{H^\nu_{1\pm}}
 + \sum_{b \in B} 
 \dfrac{
    \bra{H_{1\pm}^\nu} H_x \op{b} H_x \ket{H_{1\pm}^\nu}
 }{\varepsilon^0_{H_{1\pm}^\nu}-\varepsilon^0_b}.
\\
 \delta C_\nu + \delta M_\nu &\approx
 \sum_{b \in B} 
 \dfrac{
    |\bra{E_{1\pm}^\nu} H_0 \ket{b}|^2
 }{\varepsilon^0_{E_{1\pm}^\nu}-\varepsilon^0_b} + \cdots,
 \\
 \delta C_\nu - \delta M_\nu &\approx
 \sum_{b \in B} 
 \dfrac{
    |\bra{H_{1\pm}^\nu} H_0 \ket{b}|^2
 }{\varepsilon^0_{H_{1\pm}^\nu}-\varepsilon^0_b} + \cdots. 
\end{align}
Above, the $\delta C_\nu$ and $\delta M_\nu$ terms are higher order corrections that arise because the single QW basis states are not exact eigenstates of the triple QW $H_0$. In practice, these corrections converge slowly and we consider $\delta C_\nu$ and $\delta M_\nu$ as free parameters to adjust the band edges. For the results presented in the main text, we have neglected these corrections in Fig.~3, since it does not affect the results qualitatively. In contrast, in Fig.~5(a)-(b), near the phase transition, we have used these parameters to adjust the band edges, since it is critical to recover the correct phase.

For the inter-well couplings $v_\mu(\bm{k})$ the coefficients depend upon the eigenstates from neighboring QWs, which we label as $\nu$ and $\nu'$ next. For instance, for $\mu = LC$ we consider single QW eigenstates from $\nu = L$ and $\nu' = C$. Thus, generically, these coefficients read as
\begin{align}
 \Delta_{E1,\mu} &= 2\bra{E_{1\pm}^\nu} H_0 \ket{E_{1\pm}^{\nu'}},
 \\
 \Delta_{H1,\mu} &= 2\bra{H_{1\pm}^\nu} H_0  \ket{H_{1\pm}^{\nu'}},
 \\
 \alpha_\mu &= 2\bra{E_{1\pm}^\nu} H_x  \ket{H_{1\pm}^{\nu'}},
\end{align}
with $\Delta_{E1,\mu} = \Delta_{0,\mu} + \Delta_{z,\mu}$, and $\Delta_{H1,\mu} = \Delta_{0,\mu} - \Delta_{z,\mu}$.

\bibliography{main} 